\documentclass[9pt,twocolumn,twoside]{opticajnl}
\journal{ol} 
\setboolean{shortarticle}{true}
\usepackage{lineno}

\title{Chiral photonic topological states in Penrose quasicrystals}

\author[1]{Yingfang Zhang}
\author[2,*]{Zhihao Lan}
\author[1]{Liyazhou Hu}
\author[1]{Yiqing Shu}
\author[1]{Xun Yuan}
\author[1]{Penglai Guo}
\author[1]{Xiaoling Peng}
\author[3]{Weicheng Chen}
\author[1,*]{Jianqing Li}

\affil[1]{School of Computer Science and Engineering, Macau University of Science and Technology, Avenida Wai Long, 999078, Macau, China}
\affil[2]{Department of Electronic and Electrical Engineering, University College London, Torrington Place, London WC1E 7JE, United Kingdom}
\affil[3]{Guangdong-Hong Kong-Macao Joint Laboratory for Intelligent Micro-Nano Optoelectronic Technology, Foshan University, Foshan 528225, China}

\affil[*]{Corresponding Authors: lanzhihao7@gmail.com, jqli@must.edu.mo}

\begin{abstract}
Electromagnetic topological edge states typically are created in photonic systems with crystalline symmetry and these states emerge because of the topological feature of bulk Bloch bands in momentum space according to the bulk-edge correspondence principle.  In this work, we demonstrate the existence of chiral topological electromagnetic edge states in Penrose-tiled photonic quasicrystals made of magneto-optical materials, without relying on the concept of bulk Bloch bands in momentum space. Despite the absence of bulk Bloch bands, which naturally defiles the conventional definition of topological invariants in momentum space characterizing these states, such as the Chern number, we show that some bandgaps in these photonic quasicrystals still could host unidirectional topological electromagnetic edge states immune to  backscattering in both cylinders-in-air and holes-in-slab configurations. Employing a real-space topological invariant based on the Bott index, our calculations reveal that the bandgaps hosting these  chiral topological edge states possess a nontrivial Bott index of $\pm 1$, depending on the direction of the external magnetic field. Our work opens the door to the study of topological states in photonic quasicrystals. 
\end{abstract}

\setboolean{displaycopyright}{true}

\begin{document}

\maketitle

\par Motivated by the promise of robust device performances, topological photonics has attracted great attention recently \cite{review_RMP19,review_RiP22}. Up to now, most works in this area have considered photonic systems with crystal structures and the topological properties of the emergent edge states could be understood according to the bulk-edge correspondence principle \cite{Silveirinha_BulkEdgePRX19}. In such photonic crystal systems, the topology of bulk Bloch bands in momentum space could be characterized by topological invariants and topological features of the emergent edge states at the interface between two photonic crystal domains are determined by the difference of the topological invariants across the domain-wall interface. For example, Haldane and Raghu in 2008 \cite{Haldane08PRL} proposed to mimic the chiral edge states of electrons in the quantum Hall effect using magneto-optic photonic crystals and soon afterwards, such unidirectional backscattering-immune topological electromagnetic states were observed in experiments \cite{QH09Nature_exp}. Such photonic quantum Hall states characterized by the Chern number have found many interesting applications \cite{LuMagRev22FM}, such as topological slow-light \cite{Yang13APL_slow,Chen20OL_slow}, topological lasing \cite{Bahari17Science_lasing}, topological nonlinear frequency mixing \cite{LanPRB20_nonlinear} and topological large-area waveguide states \cite{Wang21PRL_large_area}.

\par Besides crystal structures, photonic systems with building blocks arranged in quasicrystal structures, so called photonic quasicrystals, have also received considerable attention \cite{Vardeny13NatPhot_review}. Due to the lack of exact periodicity and translational symmetry, photonic quasicrystals are not restricted to the two-, three-, four-, and six-fold rotational symmetry of crystal structures and can exhibit arbitrary rotational symmetry, such as, five- \cite{Luo14OME_5fold}, eight- \cite{Chan98PRL_8fold,Ricciardi11PRB_8fold}, ten- \cite{Notomi04PRL_10fold,Villa05PRL_10fold} and twelve-fold \cite{Zoorob00Nature_12fold,Feng05PRL_12fold,Dong15PRL_12fold} rotational symmetry. Similar to their periodic counterparts, photonic quasicrystals can also show interesting functionalities, such as bandgap formation \cite{Chan98PRL_8fold,Zoorob00Nature_12fold},  lasing action \cite{Notomi04PRL_10fold}, negative refraction \cite{Feng05PRL_12fold}, zero refractive index \cite{Dong15PRL_12fold}, and nonlinear effects \cite{Lifshitz05PRL_nonlinear,Bratfalean05OL_nonlinear}. While the study of topological states has been extended to quasicrystals in condensed matter physics recently \cite{Fan22FP_review}, it remains largely unexplored in the realm of photonics. As photonic quasicrystals can have bandgaps \cite{Chan98PRL_8fold,Zoorob00Nature_12fold}, it is natural to ask whether nontrivial bandgaps that host topologically protected edge states could be created in photonic quasicrystals.

\par In this work, we demonstrate that topological bandgaps as well as chiral topological edge states immune to backscattering could be created in Penrose photonic quasicrystals made of magneto-optical materials under external magnetic fields. To show the generality of this phenomenon, we study both dielectric cylinders in air and air holes in dielectric slab configurations and demonstrate that both configurations can host chiral topological edge states. As many important concepts, such as Bloch theorem, band structure and Brillouin zone, are not valid for photonic quasicrystals, conventional methods based on topological band theory defined in momentum space of crystal structures, such as Berry curvature and Chern number \cite{Zhao20OE_chern}, could not be used to characterize the topological states in photonic quasicrystals. Fortunately, real-space based approaches for characterizing topological states have also been developed in the literature \cite{Loring19arxiv_bott, Cerjan22NP_operator} and herein, we employ a real-space based topological invariant, called the Bott index \cite{Loring19arxiv_bott}, to explicitly verify the topological nature of the bandgaps in the proposed setups. Our calculations show that the bandgaps supporting such chiral topological edge sates host a nontrivial Bott index of $\pm 1$, depending on the direction of the external magnetic field. As the structures of photonic quasicrystals can be very rich (i.e., arbitrary rotational symmetry) and their bandgaps are more isotropic than those of crystal structures, we expect that our finding of chiral topological states in photonic quasicrystals will open a new area for topological photonics in terms of both fundamental physics and practical applications.

\begin{figure}[t!]
\centering
\includegraphics[width=\linewidth]{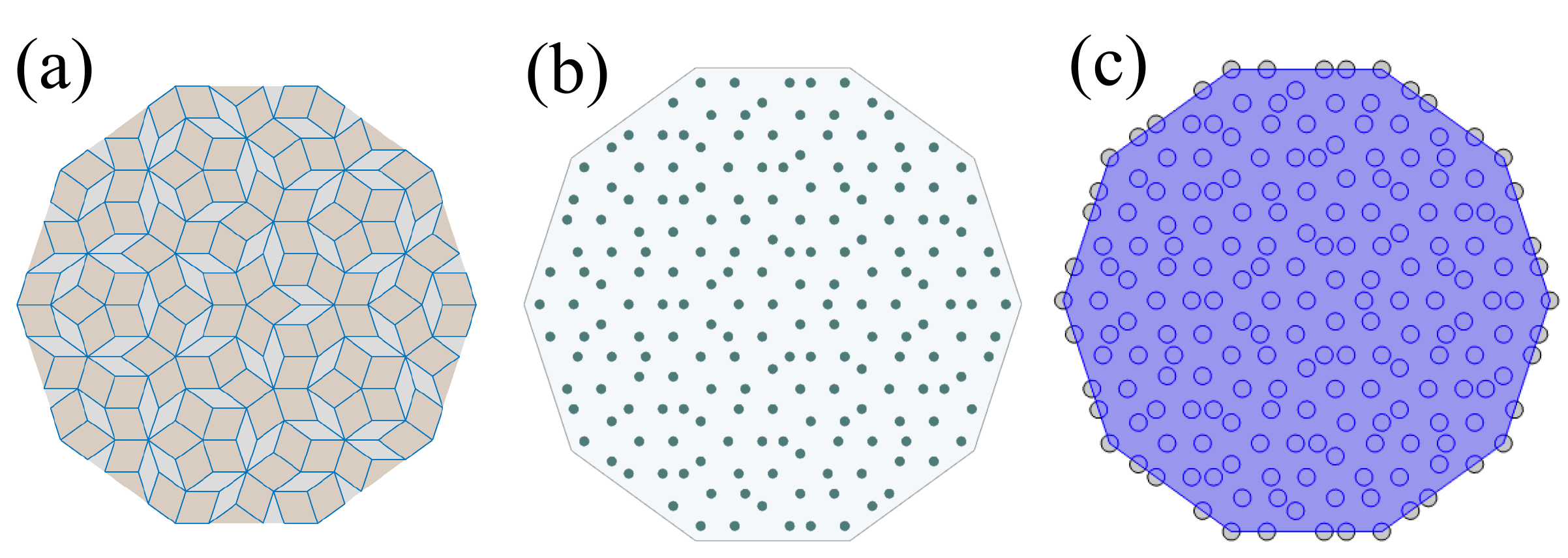}
\caption{Penrose tiling and photonic quasicrystals constructed based on the tiling. (a) A finite Penrose P3 tiling made from two rhombuses covering the area of a regular decagon. (b) Photonic quasicrystals constructed by either putting dielectric cylinders in air or etching air holes in a dielectric slab at each vertex of the tiling. (c) The structure where periodic boundary conditions could be applied at opposite edges of the computational domain indicated by the blue decagon.}
\label{fig:fig1}
\end{figure}

We consider photonic quasicrystals constructed based on the Penrose P3 tiling, which is made from two kinds of rhombuses, see Fig.\ref{fig:fig1}(a). The two kinds of rhombuses have equal sides with length $a$ and different angles. In specific, the four angles of the thin rhombus are 36, 144, 36, and 144 degrees and this rhombus consists of two acute Robinson triangles with an apex angle of 36 degrees if bisected along its short diagonal. On the other hand, the four angles of the thick rhombus are 72, 108, 72, and 108 degrees and if bisected along its long diagonal, the think rhombus can be considered to consist of two obtuse Robinson triangles with an apex angle of 108 degrees. The whole tiling is created based on a decomposition scheme as described in \cite{penrose_tile} and we choose a regular decagon containing 10 acute Robinson triangles as the starting structure for the decomposition. The tiling in Fig.\ref{fig:fig1}(a) is obtained after four iterations of decomposition and the whole tiling contains 191 vertices in total. 

\begin{figure}[t!]
\centering
\includegraphics[width=\linewidth]{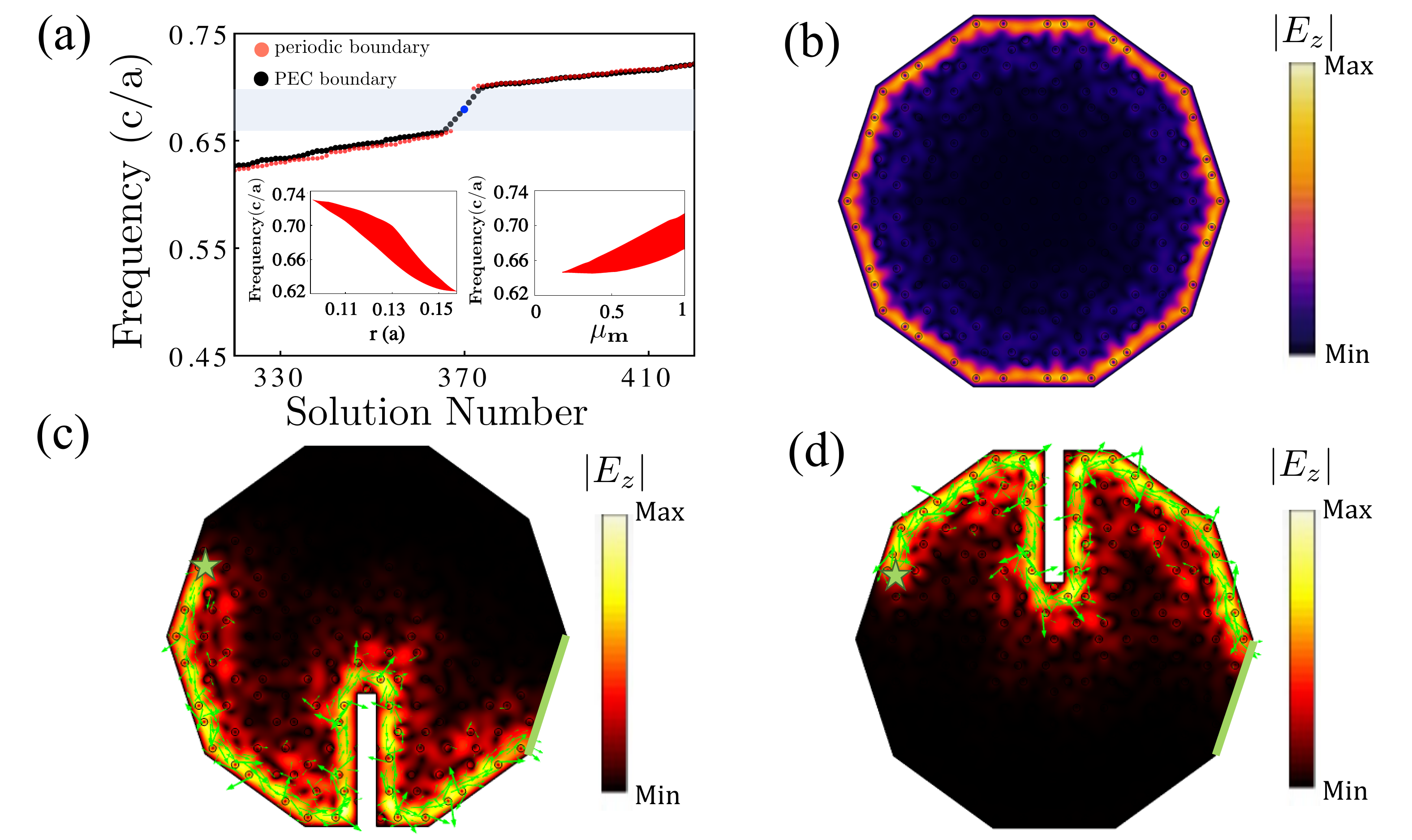}
\caption{Topological properties of Penrose photonic quasicrystals based on the cylinders-in-air configuration. (a) Topological bandgap (shaded area) and corresponding chiral edge states within the bandgap. Parameters of the cylinders are $r=0.13a$, $\epsilon=15$, $\mu_0=1$, and $\mu_{21}=-\mu_{12}=i\mu_m=i0.8$. Insets show the size of the bandgap as a function of $r$ and $\mu_m$. (b) Field profile of the edge state marked by a blue dot in (a) with PEC boundary conditions. (c) Unidirectional propagation of the edge state (green arrows are Poynting vectors) bypassing a large obstacle at $f=0.67c/a$ and $\mu_m=0.8$, where the excitation source is marked by a green star. The green edge is associated with scattering boundary condition while others are PECs. (d) Similar as (c) but with $\mu_m=-0.8$.}
\label{fig:fig2}
\end{figure}

\begin{figure}[t!]
\centering
\includegraphics[width=\linewidth]{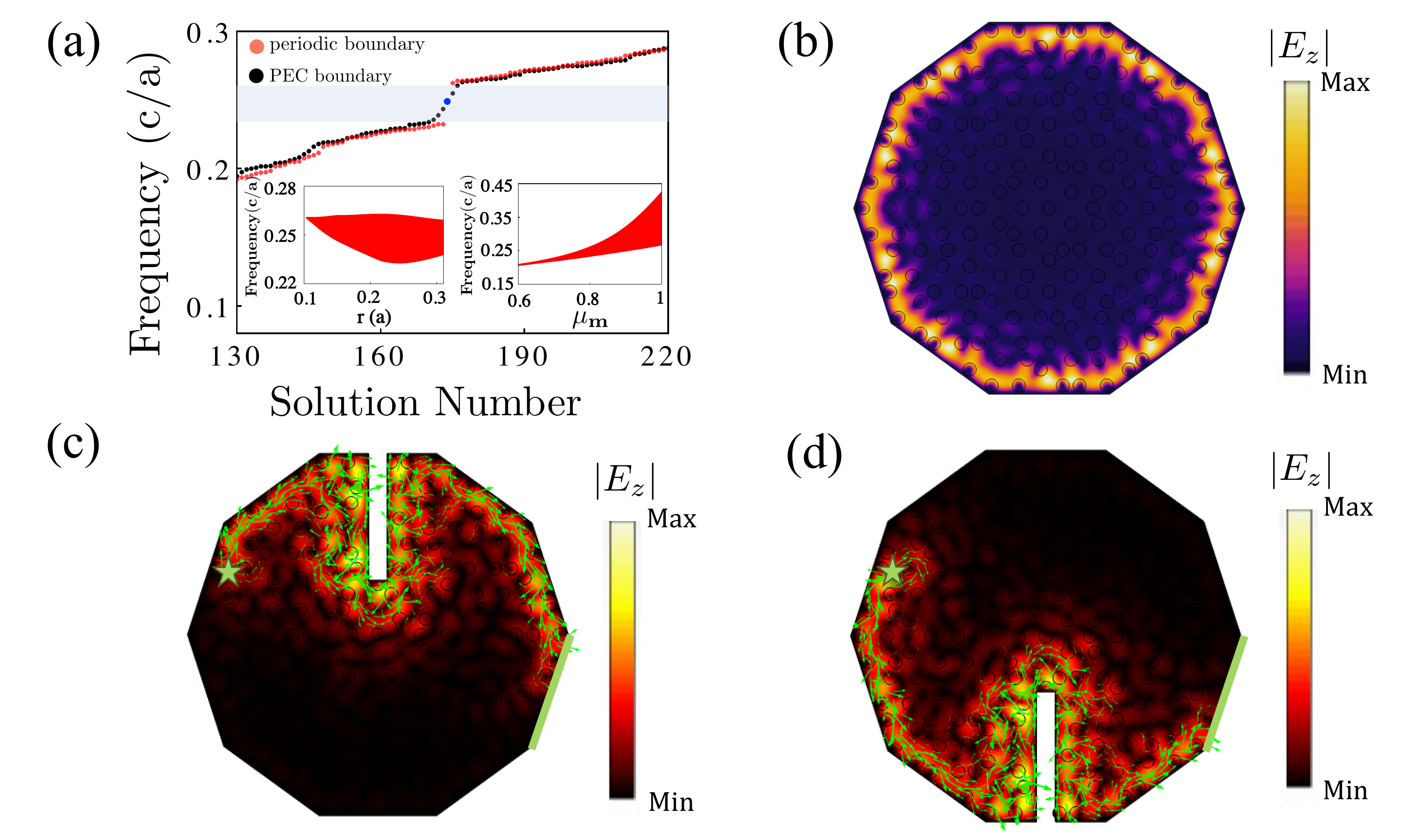}
\caption{Topological properties of Penrose photonic quasicrystals based on the holes-in-slab configuration. (a) Topological bandgap (shaded area) and corresponding chiral edge states within the bandgap. Radius of the holes is $r=0.24a$ and for the dielectric slab, $\epsilon=15$, $\mu_0=1$, and $\mu_{21}=-\mu_{12}=i\mu_m=i0.8$. Insets show the size of the bandgap as a function of $r$ and $\mu_m$. (b) Field profile of the edge state marked by a blue dot in (a) with PEC boundary conditions. (c) Unidirectional propagation of the edge state (green arrows are Poynting vectors) bypassing a large obstacle at $f=0.25c/a$ and $\mu_m=0.8$, where the excitation source is marked by a green star. The green edge is associated with scattering boundary condition while others are PECs. (d) Similar as (c) but with $\mu_m=-0.8$.}
\label{fig:fig3}
\end{figure}

A photonic quasicrystal could be constructed based on the Penrose tiling in Fig.\ref{fig:fig1}(a) in different ways, e.g., one could put photonic building blocks at the vertices of the tiling or at the center of every rhombus. In this work, we consider two kinds of photonic quasicrystal based on the vertex degrees of freedom, i.e., for the so-called cylinders-in-air configuration, dielectric cylinders with radius $r$ made of magneto-optical material are placed at the vertices of the tiling; whereas for the holes-in-slab configuration, air holes with radius $r$ are etched at the vertices of the tiling in a dielectric slab of magneto-optical material.  We will demonstrate in the following that both these two configurations can support topological bandgaps and chiral topological edge states. The regular decagonal tiling allows the study of the resulting photonic quasicrystals both in the supercell approximation with periodic boundary conditions (PBCs) and directly in a finite structure with perfect electric conductor (PEC) boundary conditions. For the supercell approximation to work, cylinders at the outer edges of the decagon are cut into halves (see the geometry in Fig.\ref{fig:fig1}(c)) and upon applying the PBCs at opposite edges of the decagon, the half cylinders within the computational domain will form complete cylinders around the boundaries, which will mimic a structure without boundary such that bandgaps of the photonic quasicrystal can be studied (see the results in Fig.\ref{fig:fig2}(a) and Fig.\ref{fig:fig3}(a)). On the other hand, we can also study the resulting photonic quasicrystal in a finite structure as in Fig.\ref{fig:fig1} (b) with PEC boundaries such that edge states could emerge within the bandgaps. Note that the supercell approximation is only a numerical technique to determine the bandgap of the photonic quasicrystals and in real experiments, geometry as in Fig.\ref{fig:fig1}(b) could be simply used to observe the topological edge states.

To demonstrate that the Penrose photonic quasicrystals can host chiral topological edge states, we first consider the cylinders-in-air configuration, which has been realized in many experiments \cite{LuMagRev22FM}. We consider yittrium iron garnet (YIG) cylinders with $r=0.13a$ and $\epsilon=15$ placed at the vertices of the  tiling, which is different from \cite{Bandres_waveguide16PRX} where coupled helical waveguides at the vertices of the tiling are used. To break the time-reversal symmetry induced by external magnetic fields, off-diagonal terms $\mu_{21}=-\mu_{12}=i\mu_m=i0.8$ are added to the permeability tensor of $\mu_0=1$ \cite{LanPRB20_nonlinear,Skirlo14PRL_largeChern}. Transverse magnetic (TM) modes with nonzero components of $\{E_z, H_x, H_y \}$ are focused in this work.

The calculated bandgap using the supercell approximation together with the edge states obtained from the structure in Fig.\ref{fig:fig1}(b) with PEC boundary conditions, are presented in Fig.\ref{fig:fig2}(a). To show these are indeed edge states, the field profile of one representative state (marked as a blue dot in Fig.\ref{fig:fig2}(a)) is given in Fig.\ref{fig:fig2}(b), from which one can see that the filed distribution is mainly confined around the edges of the system, demonstrating the edge state nature of this state. To further show that the edge states within this bandgap are chiral topological states immune to backscattering, we directly excite the edge state with a source and put a large obstacle along the propagation path as shown in Fig.\ref{fig:fig2}(c). Note that to clearly show the directionality, i.e., the propagation does not form a close loop as Fig.\ref{fig:fig2}(b), we set one edge of the decagon as scattering boundary condition whereas others as PECs. The result in Fig.\ref{fig:fig2}(c) unambiguously demonstrates the topological nature of the edge state, i.e., it propagates only along one direction and can bypass large obstacle without being backscattered. To further demonstrate that the chirality of the edge states is determined by the direction of the external magnetic field, in Fig.\ref{fig:fig2}(d) we switch the direction of the magnetic field and the result shows that the wave now also switches its propagation direction and the propagation is also immune to backscattering by obstacles. These results clearly demonstrate that photonic quasicrystals can support chiral topological edge sates, whose propagation directions are determined by the direction of the external magnetic fields and their propagations are topologically protected against backscattering by obstacles. 

To understand the properties of the topological bandgap, we study its size as a function of the radius of the cylinders and the strength of the external magnetic field, where the results are presented as two insets in Fig.\ref{fig:fig2}(a). The results show that the bandgap will disappear if the radius of the cylinders is too small or too big. On the other hand, as expected one can see that the size of the bandgap will decrease if the strength of the magnetic field becomes smaller. Nonetheless, one can still have a sizable bandgap even at moderate magnetic fields (e.g., $\mu_m=0.5$), which is convenient for the experimental observations of these chiral topological edge states in practices.

To demonstrate the generality of this topological phenomenon, we further consider Penrose photonic quasicrystals in the holes-in-slab configuration \cite{Peng22FP_holeinslab}, where air holes are etched in a YIG slab. This could be considered as the complement setup of the cylinders-in-air configuration and can have some advantages in practices as the materials are fully connected and may be more stable than the cylinders-in-air configuration. Similar as the results for the cylinders-in-air configuration, the calculated bandgap together with the edge states of the holes-in-slab configuration with hole radius of $r=0.24a$ are presented in Fig.\ref{fig:fig3}(a). Again, one can see edge states appearing within the bandgap and these states propagate unidirectionally and are immune to backscattering by large obstacles.  While the chirality (or propagation direction) of the edge states is still associated with the direction of the external magnetic fields as Figs.\ref{fig:fig3}(c, d) show, it exhibits a major difference compared with the cylinders-in-air configuration, i.e., the chirality is opposite under the same direction of the magnetic fields (see the propagations in Fig.\ref{fig:fig2}(c) and Fig.\ref{fig:fig3}(c) or Fig.\ref{fig:fig2}(d) and Fig.\ref{fig:fig3}(d)). While material dispersion of YIG is neglected in the above analysis, we find that, taking the saturation magnetization of YIG as 4$\pi M_s$ = 1780G and with an external magnetic field of about $H_0$=1000 Oe at working frequency around 10GHz \cite{QH09Nature_exp,LanPRB20_nonlinear,Wang21PRL_large_area}, the size of the topologcial bandgap is reduced by about 24$\%$  in the holes-in slab geometry by the dispersion effect whereas the chiral propagation within the topological bandgap is not affected.

While the above results demonstrate that edge states within the bandgaps are topological from a phenomenological point of view, it would be more fundamental if the topological invariant associated with the bandgap could be explicitly calculated. To to this, we adopt a real-space based topological invariant, called the Bott index \cite{Loring19arxiv_bott}, to characterize the topological bandgaps in Fig.\ref{fig:fig2}(a) and Fig.\ref{fig:fig3}(a). The real-space topological invariant of Bott index works with the phase factors \cite{Loring19arxiv_bott} of $U= \exp(2\pi i x/L)$ and  $V= \exp(2\pi i y/W)$, where $0\leq x \leq L$ and $0\leq y\leq W$ are the real space coordinates of the computational grids whereas $L$ and $W$ are the sizes of the computational domain.  Assuming P is the projector onto the space of all states below a certain frequency, the projected phase factors could then be written as  $U_1= P^{\dagger} U P $ and $V_1=P^{\dagger} V P$, and the Bott index is defined as \begin{gather}
B=\mathcal{R} \left( \frac{1}{2\pi i} \text{Tr}\left(\log (U_1V_1U_1^{\dagger}V_1^{\dagger}) \right) \right)
\end{gather}
where $\mathcal{R}$ denotes the real part of the expression. 

We briefly discuss the procedure to calculate the  Bott index in practices. Given a photonic structure, one first needs to calculate all the eigenstates (e.g., $N_f$ in total) up to a certain frequency of interest, which could be done using any eigensolver available (we use COMSOL in this work). Assuming the computational domain is discretized as $(x_m,y_m)$ with $m=1\cdots N_{xy}$, then the phase factors $U$ and $V$ are diagonal matrices of size $N_{xy}\times N_{xy} $ with $ \exp(2\pi i x_m/L)$ and $\exp(2\pi i y_m/W)$ on the diagonals. The projector $P$ is a complex matrix of size $N_{xy}\times N_f$ built from the eigen-field distributions (e.g, $E_z$ for TM modes) of all the eigenstates. Next, one could calculate the projected phase factors $U_1$ and $V_1$ of size $N_f\times N_f$ as matrix multiplications using $U, V$ and $P$. Finally, one calculates the eigenvalues of $(U_1V_1U_1^{\dagger}V_1^{\dagger})$, taking the $\log$ of them to extract the phase information, and then summing all the phases divided by $2\pi i$, which is the Bott index. As $\text{Tr}\left(\log (U_1V_1U_1^{\dagger}V_1^{\dagger}) \right) = \log \left( \det (U_1V_1U_1^{\dagger}V_1^{\dagger}) \right) $ and $\det (U_1V_1U_1^{\dagger}V_1^{\dagger}) =\det(U_1)\det^*(U_1)\det(V_1)\det^*(V_1) $, the defined Bott index is guaranteed  to be an integer. 

\begin{figure}[t!]
\centering
\includegraphics[width=\linewidth]{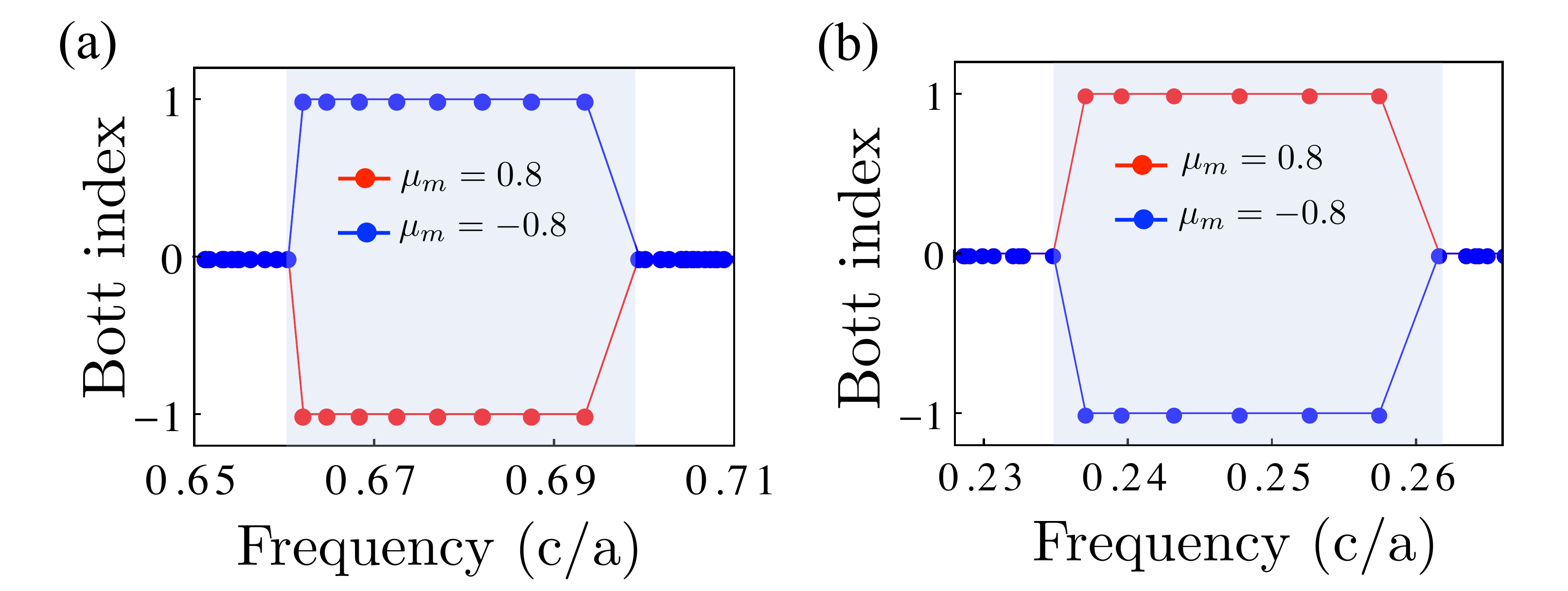}
\caption{Bott index of the edge states within the bandgap for the (a) cylinders-in-air and (b) holes-in-slab configuration. Parameters are the same as those in Fig.\ref{fig:fig2} and Fig.\ref{fig:fig3}, respectively.   }
\label{fig:fig4}
\end{figure}

The numerically calculated Bott indices for the eigenstates around the topological bandgap for both the cylinders-in-air and holes-in-slab configurations are shown in Fig.\ref{fig:fig4}, which show that the edge states within the bandgaps (shaded areas) have a nontrivial Bott index of $B= \pm 1$ depending on the direction of the external magnetic fields whereas the bulk states have a trivial Bott index of $B=0$.  More importantly, we can see that the Bott indices are opposite for the cylinders-in-air and holes-in-slab configurations at the same $\mu_m$, which is fully consistent with the chirality of propagations observed in Figs.\ref{fig:fig2}(c,d) and Figs.\ref{fig:fig3}(c,d). The unidirectional backscattering-immune propagation of the edge states together with the explicit calculations of their Bott indices unambiguously demonstrate that photonic quasicrystals constructed based on the Penrose tiling can support chiral topological edge states similar to their photonic crystal counterparts though the lack of exact periodicity.

In conclusion, we have demonstrated that photonic quasicrystals constructed based on the Penrose tiling can support chiral topological edge states despite the lack of exact periodicity and translational symmetry. This phenomenon is quite general and can exist in both cylinders-in-air and holes-in-slab configurations. Moreover, we have exploited a real-space based topological invariant -- the Bott index, to characterize these topological states and shown that they host a nontrivial Bott index of $\pm1$ depending on the direction of the applied magnetic fields.

Our work will open many interesting directions for future investigations. First, as photonic quasicrystals can exhibit arbitrary rotational symmetry \cite{Rechtsman08PRL_nfold,Gao13OE_nfold}, very rich physics could be expected. Second, whether topological states could be created in photonic quasicrystals without breaking the time-reversal symmetry, such as in the photonic quantum valley (or spin) Hall states \cite{review_RMP19,review_RiP22}, is also an interesting question and a positive sign is that Dirac cone, which is frequently used in quantum valley (or spin) Hall effects, has been demonstrated possible in photonic quasicrystals \cite{Dong15PRL_12fold}.  Furthermore, photonic quasicrystals could also be explored for studying higher-order topological states \cite{Xiong22PRApp_HighOrder,Shi22Arxiv_HighOrder} or studying topology in periodic arrays of quasicrystal units \cite{Peng22OL_Stampfli-Triangle,Xu23OL_Penrose-triangle}.  Finally, it is certainly interesting to study topological states in 3D photonic quasicrystals \cite{Man05Nature_3D,Devescovi21NC_3D} and the study of topological states in photonic quasicrystals could potentially be facilitated via advanced methods, such as inverse design \cite{Chen22_review} or automated discovery and optimization \cite{Christensen22_Automated}.

\medskip
\noindent\textbf{Funding.} This research was funded by the National Natural Science Foundation of China (No. 61827819),
and the Research Fund of Guangdong-Hong Kong-Macao Joint Laboratory for Intelligent Micro-Nano Optoelectronic Technology (No. 2020B1212030010). 

\medskip
\noindent\textbf{Disclosures.} The authors declare no conflicts of interest. 

\medskip
\noindent\textbf{Data Availability}. Data underlying the results presented in this paper are not publicly available at this time but may be obtained from the authors upon reasonable request.


\begin{thebibliography}{99}

\bibitem{review_RMP19} T. Ozawa, H. M. Price, A. Amo, N. Goldman, M. Hafezi, L. Lu, M. C. Rechtsman, D. Schuster, J. Simon, O. Zilberberg, and I. Carusotto, Rev. Mod. Phys. {\bf 91}, 015006 (2019). 

\bibitem{review_RiP22} Z. Lan, M. L. N. Chen, F. Gao, S. Zhang, and W. E. I. Sha, Rev. Phys. {\bf 9}, 100076 (2022).

\bibitem{Silveirinha_BulkEdgePRX19}M. G. Silveirinha, Phys. Rev. X {\bf 9}, 011037 (2019). 

\bibitem{Haldane08PRL}F. D. M. Haldane and S. Raghu, Phys. Rev. Lett. {\bf 100}, 013904 (2008).

\bibitem{QH09Nature_exp}Z. Wang, Y. D. Chong, J. D. Joannopoulos, and M. Soljacic, Nature {\bf 461}, 772 (2009).

\bibitem{LuMagRev22FM}X. Wang, W. Zhao, H. Zhang, S. Elshahat, and C. Lu, Front. Mater. {\bf 8}, 816877 (2022).

\bibitem{Yang13APL_slow}Y. Yang, Y. Poo, R.-X. Wu, Y. Gu, and P. Chen, Appl. Phys. Lett. {\bf 102}, 231113 (2013).

\bibitem{Chen20OL_slow}J. Chen, W. Liang, and Z.-Y. Li, Opt. Lett. {\bf 45}, 4964-4967 (2020).


\bibitem{Bahari17Science_lasing}B. Bahari, A. Ndao, F. Vallini, A. El Amili, Y. Fainman, and B. Kante, Science {\bf 358}, 636-640 (2017).

\bibitem{LanPRB20_nonlinear}Z. Lan, J.W. You, and N.C. Panoiu, Phys. Rev. B {\bf 101}, 155422 (2020).

\bibitem{Wang21PRL_large_area}M.D. Wang, R.Y. Zhang, L. Zhang, D. Wang, Q. Guo, Z.Q. Zhang, and C.T. Chan, Phys. Rev. Lett. {\bf 126}, 067401 (2021).


\bibitem{Vardeny13NatPhot_review} Z. V. Vardeny, A. Nahata, A. Agrawal, Nat. Photon. {\bf 7}, 177-187 (2013).   

\bibitem{Luo14OME_5fold} D. Luo, Q. G. Du, H. T. Dai, X. H. Zhang, and X. W. Sun, Opt. Mater. Express {\bf 4}, 1172-1177 (2014).

\bibitem{Chan98PRL_8fold}Y. S. Chan, C. T. Chan, and Z. Y. Liu, Phys. Rev. Lett. {\bf 80}, 956 (1998). 
\bibitem{Ricciardi11PRB_8fold}A. Ricciardi, M. Pisco, A. Cutolo, A. Cusano, L. O'Faolain, T. F. Krauss, G. Castaldi, and V. Galdi, Phys. Rev. {\bf B} 84, 085135 (2011). 

\bibitem{Notomi04PRL_10fold}M. Notomi, H. Suzuki, T. Tamamura, and K. Edagawa, Phys. Rev. Lett. {\bf 92}, 123906 (2004). 
\bibitem{Villa05PRL_10fold}A. D. Villa, S. Enoch, G. Tayeb, V. Pierro, V. Galdi, and F. Capolino, Phys. Rev. Lett. {\bf 94}, 183903 (2005). 

\bibitem{Zoorob00Nature_12fold} M. E. Zoorob, M. D. B. Charlton, G. J. Parker, J. J. Baumberg, and M. C. Netti, Nature {\bf 404}, 740-743 (2000). 
\bibitem{Feng05PRL_12fold} Z. Feng, X. Zhang, Y. Wang, Z.-Y. Li, B. Cheng, and D.-Z. Zhang, Phys. Rev. Lett. {\bf 94}, 247402 (2005). 
\bibitem{Dong15PRL_12fold} J.-W. Dong, M.-L. Chang, X.-Q. Huang, Z. H. Hang, Z.-C. Zhong, W.-J. Chen, Z.-Y. Huang, C. T.  Chan, Phys. Rev. Lett. {\bf 114}, 163901 (2015). 

\bibitem{Lifshitz05PRL_nonlinear}R. Lifshitz, A. Arie, A. Bahabad, Phys. Rev. Lett. {\bf 95}, 133901 (2005). 
\bibitem{Bratfalean05OL_nonlinear} R. T. Bratfalean, A. C. Peacock, N. G. R. Broderick, K. Gallo, and R. Lewen, Opt. Lett. {\bf 30},424-426 (2005). 

\bibitem{Fan22FP_review} J. Fan and H. Huang, Front. Phys. {\bf 17}, 13203 (2022). 

\bibitem{Zhao20OE_chern} R. Zhao, G.-D. Xie, M. L. N. Chen, Z. Lan, Z. Huang, and W. E. I. Sha, Opt. Express {\bf 28}, 4638-4649 (2020). 

\bibitem{Loring19arxiv_bott} T. A. Loring, arXiv:1907.11791. 
\bibitem{Cerjan22NP_operator} A. Cerjan and T. A. Loring, Nanophotonics {\bf 11},  4765-4780 (2022).

\bibitem{penrose_tile}S. Eddins, https://github.com/mathworks/penrose-tiling (2018). 

\bibitem{Bandres_waveguide16PRX} M. A. Bandres, M. C. Rechtsman, and M. Segev, Phys. Rev. X {\bf 6}, 011016 (2016).

\bibitem{Skirlo14PRL_largeChern}S. A. Skirlo, L. Lu, and M. Solja\v{c}i\'{c}, Phys. Rev. Lett. {\bf 113}, 113904 (2014).

\bibitem{Peng22FP_holeinslab} C. Peng, J. Chen, Q. Qin and Z.-Y. Li, Front. Phys. {\bf 9}, 825643 (2022).

\bibitem{Rechtsman08PRL_nfold}M. C. Rechtsman, H.-C. Jeong, P. M. Chaikin, S. Torquato, and P. J. Steinhardt, Phys. Rev. Lett. {\bf 101}, 073902 (2008).
\bibitem{Gao13OE_nfold} Y. Gao and M. Liu,  Opt. Eng. {\bf 52}, 053401 (2013). 

\bibitem{Xiong22PRApp_HighOrder} L. Xiong, Y. Zhang, Y. Liu, Y. Zheng, and X. Jiang, Phys. Rev. Applied {\bf 18}, 064089 (2022). 

\bibitem{Shi22Arxiv_HighOrder} A. Shi, J. Jiang, Y. Peng, P. Peng, J. Chen, J. Liu, arXiv:2209.05751. 

\bibitem{Peng22OL_Stampfli-Triangle}Y. Peng, E. Liu, B. Yan, J. Xie, A. Shi, P. Peng, H. Li, and J. Liu, Opt. Lett. {\bf 47}, 3011-3014 (2022).

\bibitem{Xu23OL_Penrose-triangle}Q. Xu, Y. Peng, B. Yan, A. Shi, P. Peng, J. Xie, and J. Liu, Opt. Lett. {\bf 48}, 101-104 (2023). 

\bibitem{Man05Nature_3D} W. Man, M. Megens, P. J. Steinhardt, and P. M. Chaikin, Nature {\bf 436}, 993-996 (2005).  
\bibitem{Devescovi21NC_3D}C. Devescovi, M. G.-Diez, I. Robredo, M. B. de Paz, J. L.-Alonso, B. Bradlyn, J. L. Manes, M. G. Vergniory, and A. G.-Etxarri, Nat. Commun. {\bf 12}, 7330 (2021). 

\bibitem{Chen22_review} Y. Chen, Z. Lan, Z. Su, J. Zhu, Nanophotonics  {\bf 11}, 4347-4362 (2022). 
\bibitem{Christensen22_Automated}S. Kim, T. Christensen, S. G. Johnson, and M. Solja\v{v}i\'{c}, ACS Photonics {\bf 10}, 861-874 (2023). 
  

 

\end{thebibliography}
\end{document}